\documentclass[a4paper,apx,preprintnumbers,showpacs,superscriptaddress,nofootinbib]{revtex4-2}

\usepackage{amsmath}
\usepackage{amssymb}
\usepackage{graphicx}
\usepackage{epsfig}
\usepackage{xcolor,color}
\usepackage{url}
\usepackage{bm}
\usepackage{mathrsfs}
\usepackage[utf8]{inputenc}
\usepackage{hyperref}
\usepackage{enumerate}
\usepackage{amsthm}
\usepackage{bbm}
\usepackage[normalem]{ulem}
\usepackage{upgreek}
\usepackage{tensor}
\usepackage{siunitx}
\usepackage{orcidlink}
\usepackage[caption=false]{subfig}
\usepackage{float}
\usepackage{colortbl}
\usepackage{hyperref} 
\usepackage{appendix} 
\usepackage{cleveref}
\usepackage{multirow}
\usepackage{booktabs}
\usepackage{threeparttable}
\usepackage[export]{adjustbox}
\usepackage{times}
\usepackage{commath}
 \graphicspath{{Figures/}} 
 
 \allowdisplaybreaks[3]
\hypersetup{colorlinks=true,linkcolor=blue,anchorcolor=blue,citecolor=blue}  
\definecolor{colour1}{HTML}{0571b0} 
\definecolor{colour2}{HTML}{92c5de} 
\definecolor{colour3}{HTML}{f4a582} 
\definecolor{colour4}{HTML}{ca0020} 
\definecolor{colour5}{HTML}{fe4a49} 
\definecolor{colour6}{HTML}{2d3092} 

\hypersetup{colorlinks=true, linkcolor=colour6, citecolor=colour6,
filecolor=colour6, urlcolor=colour6}

\theoremstyle{plain}

\DeclareMathOperator{\arcsinh}{arcsinh}
\DeclareMathOperator{\arccosh}{arccosh}

\newcommand{\bea}{\begin{eqnarray*}}
\newcommand{\eea}{\end{eqnarray*}}
\newcommand{\bean}{\begin{eqnarray}}
\newcommand{\eean}{\end{eqnarray}}

\newcommand{\be}{\begin{equation}}
\newcommand{\ee}{\end{equation}}

\newcommand\beq{\begin{equation}}
\newcommand\eeq{\end{equation}}
\def\bea{\begin{eqnarray}}
\def\eea{\end{eqnarray}}

\begin{document}

\title{Dictionary revision for mapping boundary data to bound states}

\author{Jiliang {Jing},\orcidlink{0000-0002-2803-7900}\footnote{Corresponding author: jljing@hunnu.edu.cn} }
 \affiliation{Department of Physics, Key Laboratory of Low Dimensional Quantum Structures and Quantum Control of Ministry of Education, and Synergetic Innovation
Center for Quantum Effects and Applications, Hunan Normal
University, Changsha, Hunan 410081, P. R. China}
\affiliation{Center for Gravitation and Cosmology, College of Physical Science and Technology, Yangzhou University, Yangzhou 225009, P. R. China}

\if 
\author{Jieci Wang\footnote{ jcwang@hunnu.edu.cn}} 
\affiliation{Department
of Physics, Key Laboratory of Low Dimensional Quantum Structures and
Quantum Control of Ministry of Education, and Synergetic Innovation
Center for Quantum Effects and Applications, Hunan Normal
University, Changsha, Hunan 410081, P. R. China}
\fi


\begin{abstract}

The correspondence between gravitational observables derived from scattering processes and adiabatic invariants in bound orbits, within the framework of the Post-Minkowskian (PM) expansion, has garnered significant attention in the study of bound orbital systems. However, the existing dictionary for this correspondence, characterized by the transformation $\beta\rightarrow i \beta$ and $b\rightarrow \pm i |b|$ with $\beta=arccosh \gamma$, produces complex-valued quantities of bound orbits  in 4PM calculations. 
Furthermore, the classical theory suggests that the changes in the scattering and  precession angles should exhibit a consistent trend with the variation in distance between the two objects. However, when the original dictionary is applied to 2PM case, the results show an opposite trend. These results contradict fundamental physical principles, thereby highlighting deficiencies in the existing dictionary. Our research identifies a critical issue: the Fourier transform of the scattering amplitude incorporates a factor of $(p_\infty^2)^{-n/2}$. This factor introduces singularities at $p_\infty^2 = 0$, thereby rendering the original dictionary become ineffective, as it assumes the possibility of connecting both scattering states and bound states at the singular point. We propose a rigorous modification by employing Hawking's method for black hole radiation, specifically analytical continuing  $p_\infty^2 \rightarrow p_\infty^2 e^{-i \pi}$. 
We also evaluate the new dictionary by comparing the binding energy calculated using effective one-body theory with numerical relativity simulation data from the SXS collaboration. Our findings indicate a remarkable agreement between the two sets of results. This revised dictionary enhances the applicability of gravitational observables derived from scattering processes to bound orbits and effectively fulfills the objectives envisioned by the pioneers who proposed this correspondence.

\end{abstract}


\maketitle

 \section{Introduction}
 
The PM expansion is particularly effective for characterizing scattering systems, and research in this area has made significant advancements. 
Pertinent  results with a 4PM accuracy have been obtained for both conservative dynamics ~\cite{Bern:2019nnu,Kalin:2020fhe,Bjerrum-Bohr:2021din,Bern:2021yeh,Dlapa:2021vgp} and radiation-reacted dynamics~\cite{Damour:2020tta,DiVecchia:2021ndb,Cho:2021arx,DiVecchia:2021bdo,Herrmann:2021tct,Bini:2021gat,Bini:2021qvf,Manohar:2022dea,Dlapa:2022lmu}. Furthermore, PM results have been expanded to incorporate considerations of spin  \cite{Bini:2017xzy,Vines:2017hyw,Bini:2018ywr,Vines:2018gqi,Guevara:2019fsj,Kalin:2019inp,Kosmopoulos:2021zoq,Aoude:2022thd,Jakobsen:2022fcj,Bern:2020buy,Bern:2022kto,FebresCordero:2022jts} and tidal effects \cite{Bini:2020flp,Bern:2020uwk,Cheung:2020sdj,Kalin:2020lmz}. Notably, using the worldline effective field theory  approach, along with the methodology of diﬀerential equations and integration by regions, Dlapa et al. \cite{Dlapa:2022lmu} obtained the total impulse in the scattering of non-spinning binaries in general relativity at the fourth PM order, including linear, nonlinear, and hereditary radiation-reaction effects.
Damour and Rettegno \cite{Damour:2022ybd} found that a reformulation of PM information in terms of effective-one-body (EOB) radial potentials leads to remarkable agreement with numerical relativity data, especially when using the radiation-reacted 4PM information. Recent significant advancements in PM theory have generated substantial interest in the development of waveform models that effectively integrate information from various perturbative approaches. This is particularly crucial for addressing the accuracy challenges presented by upcoming detectors, including the Einstein Telescope and Cosmic Explorer, as well as space-based detectors such as LISA, TianQin, and Taiji. 

\begin{figure}[htbp]
  \centering 
  \includegraphics[height=4.5cm,width=8.5cm]{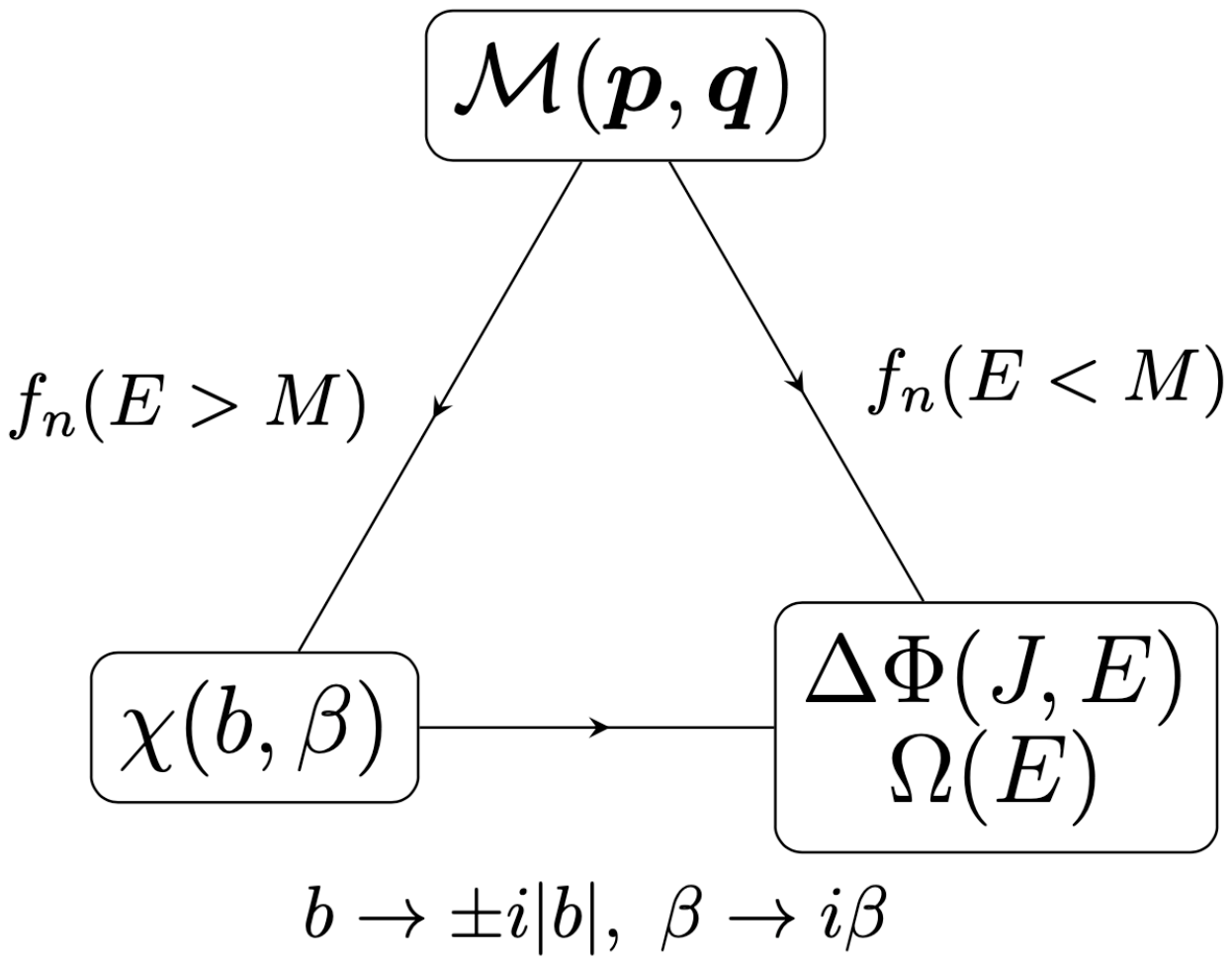}
\caption{\label{fig} The dictionary presented in reference \cite{Kalin:2019rwq}.}
\label{Dic}
\end{figure}

To establish a correspondence between gravitational observables in scattering processes and adiabatic invariants in bound orbits, K\"alin and Porto \cite{Kalin:2019rwq} introduced a conceptual framework, referred to as a dictionary, whereby the impact parameter  $b$ is transformed to $\pm i |b|$ and $\beta$ is transformed to $i \beta$ with $\beta = \arccosh \gamma$ (see fig. \ref{Dic} for detail). 
In Ref. \cite{Kalin:2019inp}, they elaborated further on this holographic mapping, which relates the periastron advance to the scattering angle.  Porto \cite{Porto:2016pyg} reviewed the effective field theory approach to gravitational dynamics. 
The primary innovation of these work lie in its ability to transform complex and intractable calculations into straightforward correspondences, thereby enhancing the study of bound orbits, particularly concerning the dynamics of binary systems in gravitational contexts. This methodology has been widely adopted in research focused on bound orbital systems. 
K\"alin et al. \cite{Kalin:2020fhe} derived the conservative dynamics of non-spinning binaries to third PM order using the eﬀective field theory approach introduced in \cite{Kalin:2020mvi}, along with the Boundary-to-Bound (B2B) dictionary developed in \cite{Kalin:2019inp,Kalin:2019rwq}.
Building upon this framework, Ref. \cite{Kalin:2020lmz} computed the next-to-leading order PM tidal eﬀects within the conservative dynamics of compact binary systems. 
Ref. \cite{Liu:2021zxr} presented a method for constructing the radial action for elliptic-like orbits using the B2B correspondence, deriving the aligned-spin radial action from the resulting scattering data, and calculated the periastron advance and binding energy for circular orbits.
Ref. \cite{Dlapa:2021npj} provided an analysis of the contribution from potential interactions to the dynamics of non-spinning binaries up to 4PM order, achieved by computing the scattering angle to $O(G^4)$ using the eﬀective field theory approach, and deriving the bound radial action through analytic continuation.
Aiming to compute bound waveforms from scattering amplitudes at 3PM, Ref.  \cite{Adamo:2024oxy}  investigated gravitational two-body dynamics utilizing the Schwinger-Dyson equations and Bethe-Salpeter recursion.  Ref. \cite{Wilson-Gerow:2025xhr} examined the conservative dynamics of spinless compact objects  within a gravitational framework that includes a metric and an arbitrary number of scalar fields up to 3PM. The result encodes the fully relativistic dynamics of the compact objects. Ref. \cite{Dlapa:2024cje} analyzed the local-in-time conservative binary dynamics at fourth PM order, while the paper \cite{Buonanno:2024vkx}  presented the derivation of the first spinning EOB Hamiltonian incorporating PM results up to three-loop order. The model accounts for the complete hyperbolic motion, encompassing nonlocal-in-time tails. Notably, This methodology has been shown to enhance gravitational wave models for binary signals ~\cite{Chiaramello:2020ehz,Nagar:2021gss,Placidi:2021rkh,Khalil:2021txt,Ramos-Buades:2021adz,Nagar:2021xnh,Khalil:2022ylj,Buonanno:2024byg,Jing2019,Jing,Jing1,JingXZ,Jing3}.

Recently, reference \cite{Buonanno:2024byg} utilizes 4PM scattering angle to develop a comprehensive inspiral-merger-ringdown waveform model for non-precessing spinning black holes within the EOB formalism. However, in this study, to prevent the emergence of complex-valued results that violate physical principles by using the dictionary established in Reference \cite{Kalin:2019rwq},  certain terms in the correspondence results were modified. Specifically, the modifications included substitutions such as $ Li_2\Big(\sqrt{\frac{\gamma-1}{\gamma+1}}\Big) \rightarrow Ti_2 \Big(\sqrt{\frac{1-\gamma}{\gamma+1}}\Big) $ and $ \log (\gamma^2-1) \rightarrow \log(u) $, where $  u = M/r $. Additionally, terms that posed challenges for handling, such as $ \Big(\frac{3 \sqrt{\gamma ^2-1} h_7}{8 (\gamma -1)^2 (\gamma +1)^3}\Big) \text{Li}_2\left(\frac{\gamma -1}{\gamma +1}\right) $, were omitted.

\begin{figure}[htbp]
  \centering 
  \includegraphics[height=7.5cm,width=9.5cm]{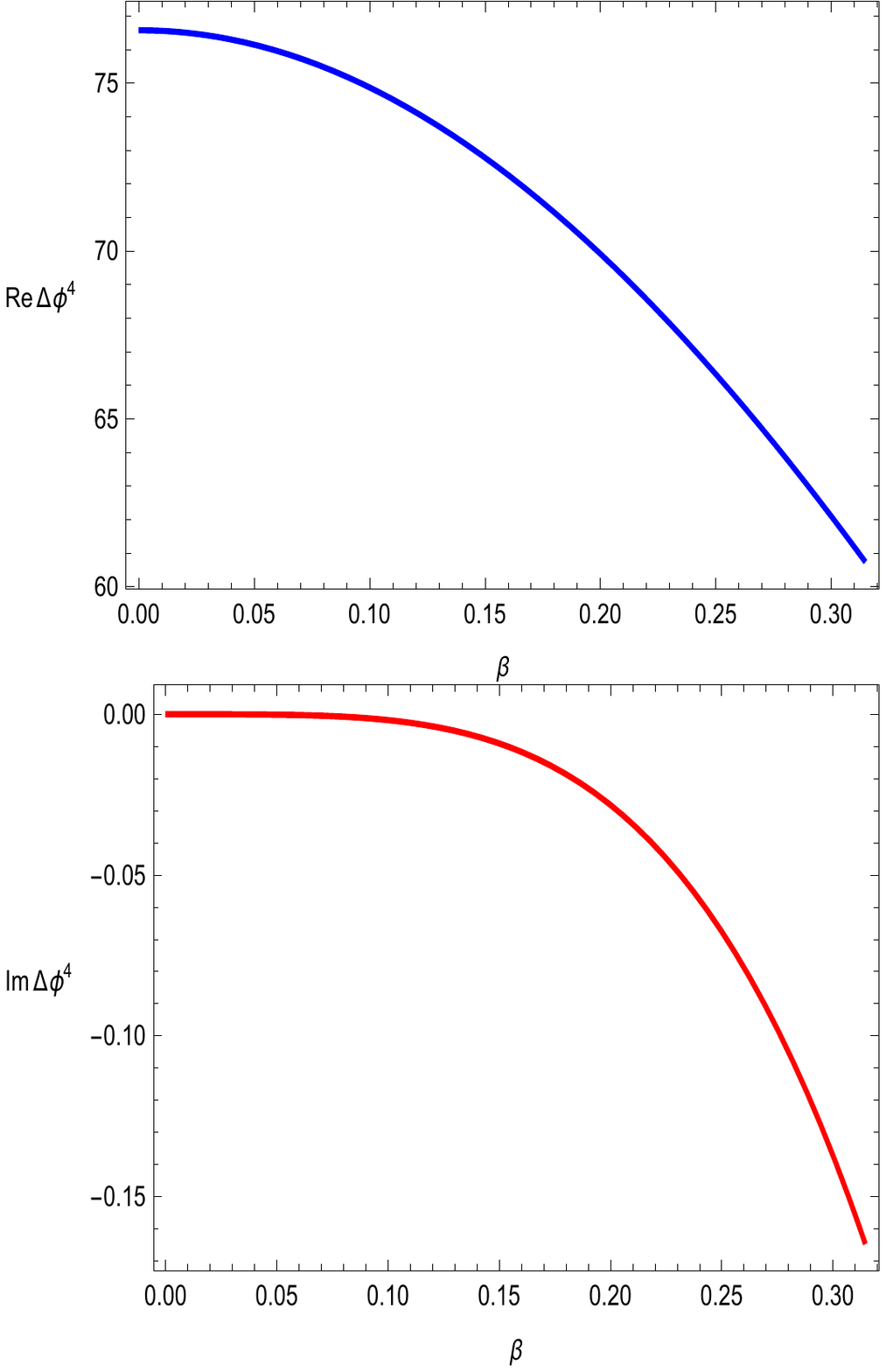}
\caption{\label{fig}  The upper panel is for $Re[\Delta \Phi_j^{4}]$, and the lower one for $Im[\Delta \Phi_j^{4}]$.}
\label{fig1}
\end{figure}

Utilizing the relationship between the scattering angle and the periastron advance, represented as $ \Delta \Phi_j^{2 n} = 2 \chi_j^{(2 n)} $ \cite{Porto:2016pyg}, along with the dictionary provided by K\"alin and Porto, we can demonstrate that the 4PM  periastron advance $ \Delta \Phi^{4} $ is complex-valued, which violates fundamental physical principles. For further details, please refer to figure \ref{fig1}.

\begin{figure}[htbp]
  \centering 
  \includegraphics[height=5.5cm,width=9.5cm]{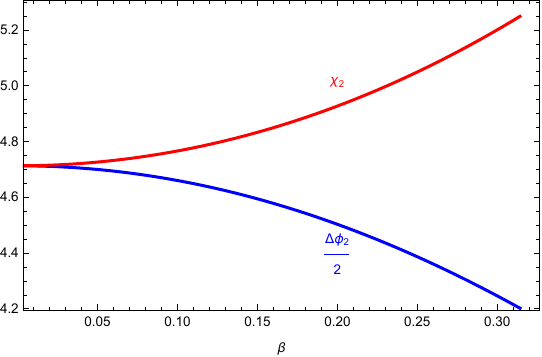}
\caption{\label{fig}  When the original dictionary is applied to 2PM, the precession angle actually decreases as the distance between the two objects diminishes.}
\label{figa}
\end{figure}

According to classical theory, the variations of the scattering angle and the precession angle are expected to follow consistent trends as the distance between two objects changes. However, when the original dictionary is applied to the 2PM case, the results indicate that as the distance between the two objects decreases, the scattering angle increases, while the precession angle displays a contrary tendency.
(see fig. \ref{figa} for detial). 

These results obtained using the original dictionary clearly contradict fundamental physical principles, thereby underscoring the deficiencies present in the existing dictionary.

In this paper, we will identify the root causes of this issue and address them comprehensively. 

This paper is organized as follows. In section~\ref{secFT}, one shows that the Fourier transform of the scattering amplitude includes a factor of $(p_\infty^2)^{-n/2}$, which signifies a singularity at $p_\infty^2 = 0$.  In section~\ref{secEP},  one extends the applicability of gravitational observables from scattering processes to bound orbits. The conclusions are given in section~\ref{Sum}.

\section{Fourier transform of scattering amplitude signifies a singularity at $p_\infty^2 = 0$\label{secFT}}

In the subsequent studies, we will first identify the root causes of the issue and then address them comprehensively. To uncover the underlying cause, we will conduct a concise review of the calculation process for the scattering angle.

Bern et al. \cite{PhysRevLett.122.201603,Bern20192} demonstrated that the conservative Hamiltonian for a relativistic massive spinless binary system, consisting of masses $ m_1 $ and $ m_2 $, is described by 
\begin{eqnarray}
H(\vec{p},\vec{r})=\sqrt{|\vec{p}|^{\;2}+m_1^2}+\sqrt{|\vec{p}|^{\;2}+m_2^2}+\sum_{i=1}^{\infty}c_i\Big(\frac{G}{|\vec{r}|}\Big)^i,
\end{eqnarray} 
where $ \vec{r} $ is the distance vector between the particles, $ \vec{p} $ represents the momenta, $ i $ labels PM orders, and $ c_i=c_i(p_\infty^2)\, (i=1,2,3) $ can be found in Refs. \cite{PhysRevLett.122.201603,Bern20192}. It is noteworthy that there are two conserved quantities: the energy $ \mathcal{E}=-p_t $ and the angular momentum $ J=p_{\phi} $. Consequently, the reduced action may be expressed as $
S=-\mathcal{E}t+J\phi+S_r(r,\mathcal{E},J).
$ Utilizing the Hamilton-Jacobi equation $
\frac{\partial S}{\partial t}+H(q,\frac{\partial S}{\partial q})=0,
$ we can derive the radial momentum $ p_r^2=(\frac{dS_r}{dr})^2 $ as a function of the radial coordinate $ r $. Up to the 4PM order, this is expressed as \cite{Bern20192, Jing3} 
\begin{align}\label{Pr2}
p_r^2 &= \frac{P_0 r^2 - J^2}{r^2} + P_1 \left(\frac{G}{r}\right) + P_2 \left(\frac{G}{r}\right)^2\nonumber \\ & + P_3 \left(\frac{G}{r}\right)^3 + P_4 \left(\frac{G}{r}\right)^4,
\end{align}
where $ P_n $ denotes the coefficients in the PM expansion derived from the following Fourier transform of the scattering amplitude \cite{Bern20192} 
\begin{align}
\widetilde{\mathcal{M}}(r, E) &= \frac{1}{2E} \int \frac{d^3 \mathbf{q}}{(2\pi)^3} \mathcal{M}(\mathbf{q}, p^2_\infty = p_\infty^2(E)) e^{-i \mathbf{q} \cdot \mathbf{r}} \nonumber \\ &= \sum_{n=1}^\infty P_n \left(\frac{G}{r}\right)^n.
\label{FT}
\end{align}
For concise representation, in the following, we will express $ P_0 $ as 
\begin{align}
P_0 &= \frac{(\mathcal{E}^2-(m_1-m_2)^2)(\mathcal{E}^2 - M^2)}{4\mathcal{E}^2} \nonumber \\ & \equiv \left(\frac{\mu}{\Gamma}\right)^2 p_\infty^2\nonumber \\ & \equiv \left(\frac{\mu}{\Gamma}\right)^2 (\gamma^2 - 1),
\end{align}
where $M = m_1 + m_2$, $\Gamma = \frac{\mathcal{E}}{M}$, and $\mu = \frac{m_1 m_2}{M}$. Here, $p_\infty^2 = (\gamma^2 - 1)$, and $\mathcal{E}$ represents the relativistic energy of the two-body system. For scattering states, it holds that $\mathcal{E} - M > 0$, while for bound states, $\mathcal{E} - M < 0$. This shows that the primary distinction between these two scenarios is that $p_\infty^2 > 0$ ($\gamma > 1$) characterizes scattering states, while $p_\infty^2 < 0$ ($\gamma < 1$) pertains to bound states. Consequently, the coefficients $ P_n $ are specified as follows
\begin{align}
P_1 &= 2M\mu^2\left(\frac{2p_\infty^2 + 1}{\Gamma}\right), \nonumber \\
P_2 &= \frac{3M^2\mu^2}{2} \left(\frac{5p_\infty^2 + 4}{\Gamma}\right), \nonumber
\\
P_3&= \frac{M^3 \mu^2 }{\Gamma } \Bigg\{\frac{17}{2}+9 p_\infty ^2+\frac{\nu}{6} \Bigg[\frac{18 \Gamma (2p_\infty ^2+1) (5 p_\infty ^2+4)}{(\Gamma +1) \left(\sqrt{p_\infty ^2+1}+1\right)}\nonumber \\ &- 4 (p_\infty ^2+1)^{3/2}-108 (p_\infty ^2+1) 
-206 \sqrt{p_\infty ^2+1}+6\Bigg]\nonumber \\ &+\frac{8 \nu \left(11+4 p_\infty ^2-4 p_\infty ^4\right) \arcsinh\sqrt{\frac{\sqrt{p_\infty ^2+1}-1}{2}}}{\sqrt{p_\infty ^2}}\Bigg\},\nonumber \\
P_4&= -\frac{9\, M^2 \mu^2\,\left(4+5\, p_\infty ^2 \right)^2}{8\, p_\infty ^2}-\frac{2\, M\, \Gamma \,\left(1+2\, p_\infty ^2 \right) P_3}{ p_\infty ^2 }\nonumber \\ &+\frac{4 \, M^2\, \mu^2\, \Gamma^2\, T_{4p}}{3\,\pi \, p_\infty ^2 }, \label{P01234}
\end{align}
with 
\begin{align}
T_{4p} &=\frac{h[61]}{16 p_\infty ^2 \Gamma ^3}+\frac{ p_\infty^4 \nu }{144 \Gamma ^3} \Bigg\{\frac{36 }{ p_\infty^4}\text{ E}\Big(\frac{\sqrt{1+ p_\infty^2 }-1}{1+\sqrt{1+ p_\infty^2 }}\Big) \text{ K}\Big(\frac{\sqrt{1+ p_\infty^2 }-1}{1+\sqrt{1+ p_\infty^2 }}\Big) h_{4}-\frac{24 \,\pi^2\, h_{5}}{ p_\infty^2 }\nonumber \\ &-\frac{126 \text{ E}\Big(\frac{\sqrt{1+ p_\infty^2 }-1}{1+\sqrt{1+ p_\infty^2 }}\Big)^2 h_{2}}{p_\infty^2 (\sqrt{1+ p_\infty^2 }-1)} 
-\frac{36}{ p_\infty^4} \text{ K}\Big(\frac{\sqrt{1+ p_\infty^2 }-1}{1+\sqrt{1+ p_\infty^2 }}\Big)^2 h_{3}+\frac{12\, h_{24} \, \text{arccosh}\sqrt{1+ p_\infty^2 }}{ p_\infty^7} \nonumber \\
&+\frac{18 \,h_{26}\,\text{arccosh}(\sqrt{1+ p_\infty^2 })^2 }{ p_\infty^8}-\frac{h_{62}}{ p_\infty^6 (1+ p_\infty^2 )^{7/2}}-\frac{48\, h_{23}\, \text{log}(1+ p_\infty^2 )}{ p_\infty^4} \nonumber \\
&-\Big(\frac{12\, h_{6} }{ p_\infty^4}+\frac{36\, h_{16}\,\text{arccosh}(\sqrt{1+ p_\infty^2 })}{ p_\infty^5}\Big)\text{log}\Big(\frac{1}{2} \Big(\sqrt{1+ p_\infty^2 }-1\Big)\Big) \nonumber \\
&-\Big(\frac{12\, h_{22} }{ p_\infty^4}+\frac{36\, h_{28} \, \text{arccosh}\Big(\sqrt{1+ p_\infty^2 }\Big)}{ p_\infty^5}\Big)\text{log}\Big(\frac{1}{2} \Big(1+\sqrt{1+ p_\infty^2 }\Big)\Big) \nonumber \\
&+\frac{72 h_{15} \text{log}(\frac{1}{2} (\sqrt{1+ p_\infty^2 }-1)) \text{log}(\frac{1}{2} (1+\sqrt{1+ p_\infty^2 }))}{ p_\infty^2 }-\frac{48 h_{29} }{ p_\infty^2 } \text{Li}_2(\frac{1-\sqrt{1+ p_\infty^2 }}{1+\sqrt{1+ p_\infty^2 }}) \nonumber \\
&-\frac{288\, h_{27}\, \text{log}\Big(\frac{1}{2} \Big(1+\sqrt{1+ p_\infty^2 }\Big)\Big)^2}{ p_\infty^2 }-\frac{576\, h_{7} \, \sqrt{ p_\infty^2 } \, \text{Li}_2\Big(\sqrt{\frac{\sqrt{1+ p_\infty^2 }-1}{1+\sqrt{1+ p_\infty^2 }}}\Big)}{ p_\infty^4 (1+\sqrt{1+ p_\infty^2 })} \nonumber \\
&+\frac{72\, (2\, \sqrt{ p_\infty^2 }\, h_{7}- p_\infty^2 (1+\sqrt{1+ p_\infty^2 })\, h_{30}) \text{Li}_2\Big(\frac{\sqrt{1+ p_\infty^2 }-1}{1+\sqrt{1+ p_\infty^2 }}\Big)}{ p_\infty^4 (1+\sqrt{1+ p_\infty^2 })}\Bigg\}.  
\end{align}
where the quantities $h_i$ can be, for clarity, found in Ref. \cite{JingNew}.

Then, using the definition $
\chi=-\pi+2 J \int_{r_{\rm min}}^{\infty}\frac{d r}{r^2\sqrt{p_r^2}}, \label{chireal}
$ where the minimum distance $r_\text{min}$ is determined by $p_r(r_{min})=0$, one can derive the scattering angle up to the 4PM order, which is given by 
\begin{eqnarray}
\chi=\chi_1\frac{G}{J}+\chi_2\left(\frac{G}{J}\right)^2+\chi_3 \left(\frac{G}{J}\right)^3+\chi_4 \left(\frac{G}{J}\right)^4 \;,
\end{eqnarray}
with 
\begin{align}
 &\chi_1=\frac{P_1}{\sqrt{P_0}}, \nonumber \\
 &\chi_2=\frac{\pi }{2 } P_2, \nonumber \\ 
 & \chi_3= -\frac{ P_1^3-12 P_0 P_1 P_2-24 P_0^2 P_3}{12 P_0^{3/2}}, \nonumber \\
& \chi_4=\frac{3 \pi }{8}(P_2^2+2 P_1 P_3+2 P_0 P_4). \label{4pmchi}
\end{align}

It is evident that the Fourier transform (\ref{FT}) includes a factor of $(p_\infty^2)^{-n/2}$, which signifies a singularity at $p_\infty^2 = 0$ in the radial momentum (\ref{Pr2}). Therefore, these related results are valid only for scattering states $p_\infty^2 > 0$. To extend their applicability to bound states $p_\infty^2 < 0$, one must implement analytic continuation.

\begin{figure}[htbp]
  \centering 
  \includegraphics[height=3.8cm,width=8.2cm]{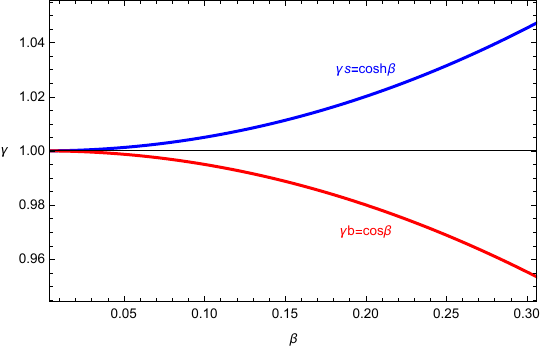}
\caption{The upper curve represents scattering states, denoted as $\gamma_s = \cosh \beta$, while the lower curve corresponds to bound states, $\gamma_b = \cos \beta$. Both connect at $\beta=0$.}
\label{fig2}
\end{figure}

The original dictionary \cite{Kalin:2019rwq} is derived from results obtained through the 3PM expansion, in which no singularity is present (see Eq. (\ref{Pr2})). This is seem reasonable, as scattering states ($\gamma_s = \cosh \beta > 1$) can be analytically continued into bound states ($\gamma_b = \cos \beta < 1$), with both states interconnected at $\beta = 0$ (see Figure \ref{fig2}).

It is crucial to emphasize that the entries in the original dictionary—namely, ($\gamma = \cosh \beta$) and ($\beta \rightarrow i \beta$)—have lost their validity due to the singularity at $p_\infty^2 = 0$ in radial momentum up to the 4PM order. This loss arises because both scattering states and bound states cannot be connected at the singular point $p_\infty^2 = 0$, but the original dictionary suggesting that such a connection is feasible. Consequently, applying the original dictionary to calculations up to the 4PM order results in complex-valued quantities associated with bound orbits.

\section{Extend applicability of gravitational observables from scattering processes to bound orbits\label{secEP}}

How can we analytically continue the results applicable to $p_\infty^2 > 0$ into the realm of $p_\infty^2 < 0$? Fortunately, the challenges encountered in mathematics align precisely with those investigated by Hawking in his research on black hole radiation \cite{Hawking75}. By employing analytic continuation, Hawking \cite{Hawking75} extended positive-frequency coefficients of the Bogoliubov transformation ($\alpha_{\omega \omega'}$) around the singularity ($\omega' = 0$) into the negative-frequency domain $\alpha_{\omega (-\omega')}$.
To achieve this analytic continuation of results applicable to positive values ($p_\infty^2 > 0$) around the singularity ($p_\infty^2 = 0$) into the negative values ($p_\infty^2 < 0$), we note that $p_r^2$ is obtained from the Fourier transform of a function $\mathcal{M}(\mathbf{q}, p^2_\infty = p_\infty^2(E))$, which is analytic for $p_\infty^2 > 0$. Consequently, $p_r^2$ is analytic in the upper half of the $p^2_\infty$ plane, necessitating the selection of branch cuts in the lower half-plane. This is accomplished by substituting $ p_\infty^2 $ with $ p_\infty^2 e^{-i \pi} $, which also incorporates an additional entry in the original dictionary: $ b = J/p_\infty \rightarrow \pm i |b| $. Thus, we extend the applicability of gravitational observables from scattering processes—initially valid for $p_\infty^2 > 0$—to bound orbits where $p_\infty^2 < 0$. 

\begin{figure}[htbp]
  \centering 
  \includegraphics[height=5.5cm,width=9.5cm]{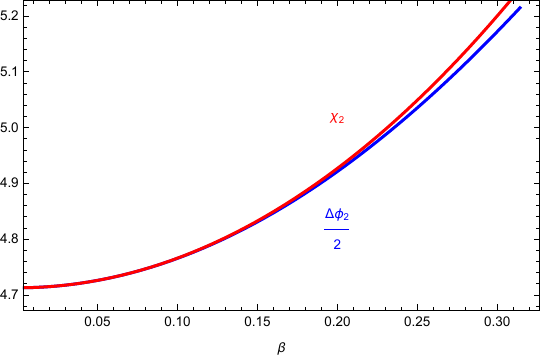}
  \includegraphics[height=5.5cm,width=9.5cm]{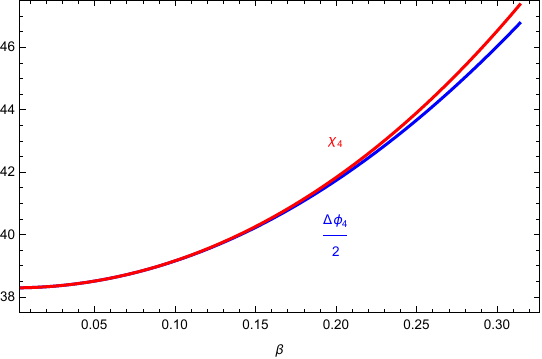}
  \caption{\label{fig}  The figure shows that by using the new dictionary, as two objects approach one another, both the scattering angle and precession angle increase.}
\label{figb}
\end{figure}

Now we evaluate the new dictionary begins with the dictionary is applied to 2PM and 4PM cases. The Figs. \ref{figb} show that as two objects approach one another, both the scattering angle and precession angle increase. 

\begin{figure}[htbp]
  \centering 
  \includegraphics[height=8.5cm,width=9.5cm]{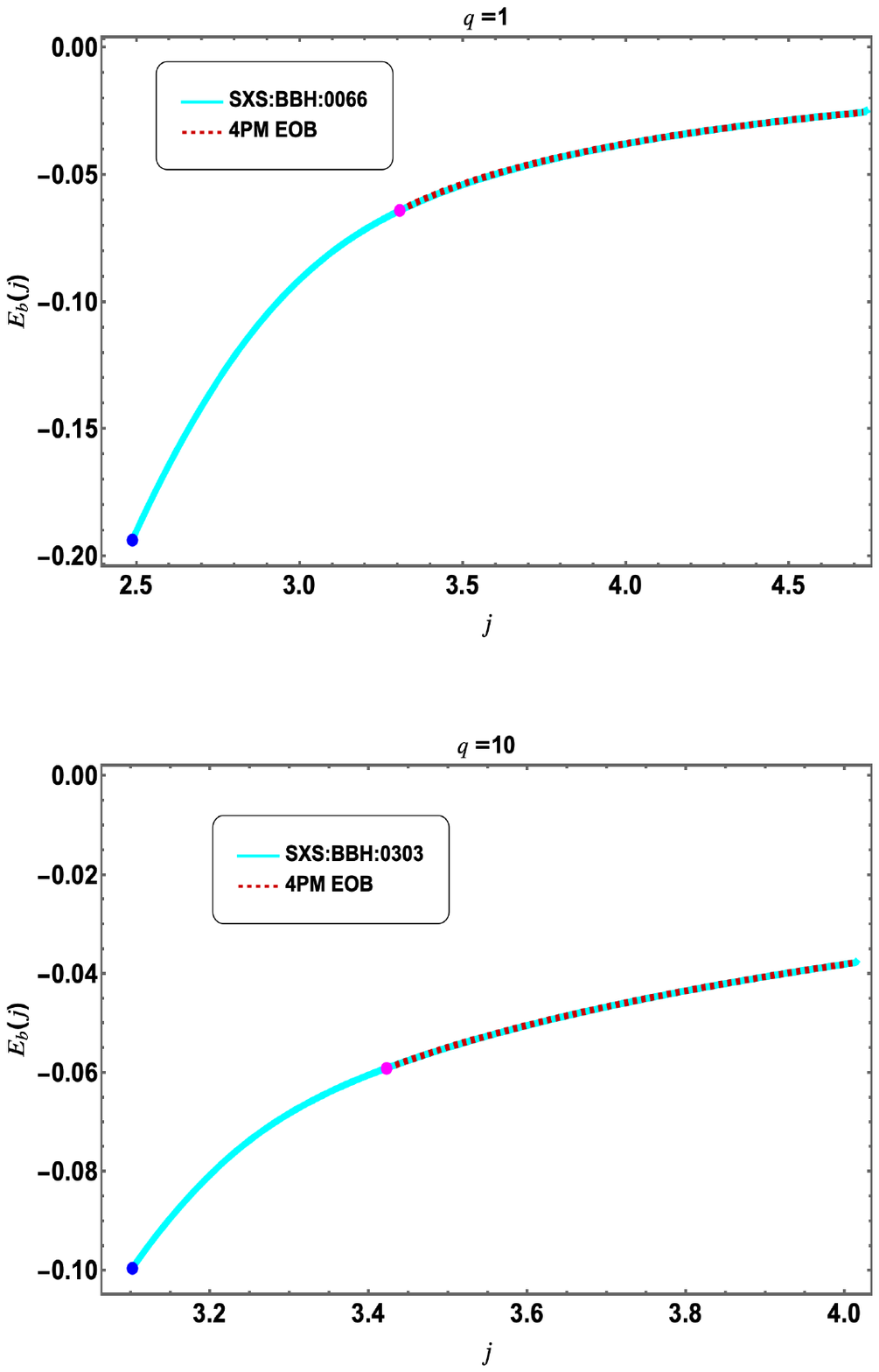}
	\caption{The relation plots of $E_{b}(j)$ with the mass ratio of $q=1,~10$, where pink solid markers denotes the innermost stable circular orbit calculated by EOB to 4PM order, and black solid markers represents the final state of the black hole in NR simulations from the SXS collaboration.}
		\label{fig:sxsbbh2516}
	\end{figure}

Furthermore, by comparing the binding energy $E_{b}(j)$ (where $j$ denotes the orbital angular momentum) calculated using EOB theory up to 4PM order \cite{Jing,Jing1,JingXZ,Jing3} with results obtained from NR simulations since the binding energy is a crucial component in the computation of gravitational waveforms emitted by coalescing binaries  \cite{PhysRevLett.108.131101,PhysRevD.93.044046}. The NR simulation data are sourced from the SXS database, specifically from the signals labeled \texttt{SXS:BBH:0066 (q=1) and SXS:BBH:0303 (q=10)}. As illustrated in Figures \ref{fig:sxsbbh2516}, the relationship between binding energy and angular momentum concerning the innermost stable circular orbit, as derived from EOB theory, demonstrates a remarkable agreement with NR data.

\section{Summary\label{Sum}}

In summary, the correspondence between gravitational observables derived from scattering processes and adiabatic invariants in bound orbits within the context of the PM expansion has gained considerable acceptance in the study of bound orbital systems. 
However, the existing dictionary for this correspondence, characterized by the transformation $\beta\rightarrow i \beta$ and $b\rightarrow \pm i |b|$ with $\beta=arccosh \gamma$, produces complex-valued quantities of bound orbits  in 4PM calculations. 
Furthermore, classical theory posits that the variations in the scattering angle and the precession angle are should exhibit consistent trends as the distance between two objects changes. However, when the original dictionary is applied to the 2PM case, the results indicate a contrary tendency. These findings contradict fundamental physical principles, thereby underscoring deficiencies in the existing dictionary. 
Our study identifies a critical issue: the Fourier transform for the scattering amplitude includes a factor of $(p_\infty^2)^{-n/2}$ which emerges in fourth-order PM. This indicates that the results obtained from scattering amplitudes exhibit singularities when $ p_\infty^2 = 0$. This phenomenon illustrates that the original dictionary entry (e.g., $\gamma = \cosh \beta$ and $\beta \rightarrow i \beta $) becomes ineffective, as it assumes the possibility of connecting both scattering states and bound states at the singular point $\beta = 0$ or $p_\infty^2 = 0$.
To address this issue, we propose a rigorous modification to the dictionary by employing Hawking's approach to black hole radiation, which involves analytically continuing positive-frequency results around the singularity into the negative-frequency domain. By applying the analytic continuation $p_\infty^2 \rightarrow p_\infty^2 e^{-i \pi}$, we extend the applicability of gravitational observables from scattering processes—initially valid for $p_\infty^2 > 0$—to bound orbits where $p_\infty^2 < 0$. 
We evaluate the new dictionary by applying it to the 2PM and 4PM cases. The results indicate that as the two objects approach one another, both the scattering angle and the precession angle exhibit consistent trends.
Furthermore, by comparing the binding energy calculated using EOB theory with the NR data from the SXS collaboration. Our findings indicate that the binding energy-angular momentum relationship to the innermost stable circular orbit derived from EOB theory is in remarkable agreement with NR data.
This revised dictionary enhances the applicability of gravitational observables derived from scattering processes to bound orbits and effectively fulfills the objectives envisioned by the pioneers who proposed this correspondence.

\acknowledgments
{ This work was supported by the Grant of NSFC Nos. 12035005, and National Key Research and Development  Program of China No. 2020YFC2201400.}  
\bibliography{mybib}

\end{document}